# MetaCompass: Reference-guided Assembly of Metagenomes

Tu Luan[1,2,+], Victoria Cepeda[1,2,+], Bo Liu[1,2,+], Zac Bowen[3], Ujjwal Ayyangar[3], Mathieu Almeida[2], Christopher M. Hill[1,2], Sergey Koren[4], Todd J. Treangen[2], Adam Porter[1], Mihai Pop[1,2]*

[1]Department of Computer Science, University of Maryland, College Park, Maryland, USA.

[2]Center for Bioinformatics and Computational Biology, University of Maryland, College Park, Maryland, USA.

[3] Fraunhofer USA Center Mid-Atlantic, Riverdale, Maryland, USA.

[4] Genome Informatics Section, Computational and Statistical Genomics Branch, National Human Genome Research Institute, Bethesda, Maryland, USA.

+ These authors contributed equally to this work

* mpop@umd.edu, to whom correspondence should be addressed

## ABSTRACT

Metagenomic studies have primarily relied on *de novo* assembly for reconstructing genes and genomes from microbial mixtures. While reference-guided approaches have been employed in the assembly of single organisms, they have not been used in a metagenomic context. Here we describe the first effective approach for reference-guided metagenomic assembly that can complement and improve upon *de novo* metagenomic assembly methods for certain organisms. Such approaches will be increasingly useful as more genomes are sequenced and made publicly available.

## Background

Microorganisms play an important role in virtually all of the Earth's ecosystems and are critical for the health of humans [1], plants, and animals. Most microbes, however, cannot be easily grown in a laboratory [2]. The analysis of organismal DNA sequences obtained directly from an environmental sample (a field termed metagenomics), enables the study of microorganisms that are not easily cultured. Metagenomic studies have exploded in recent years due to the increased availability of inexpensive high-throughput sequencing technologies. Some examples include the MetaHIT project in Europe [3], the Human Microbiome Project (HMP) in the US [4], as well as crowd sourced projects such as American Gut [5].

The analysis of these vast amounts of data is complicated by the fact that reconstructing large genomic segments from metagenomic reads is a formidable computational challenge. Even for single organisms, the assembly of genome sequences from short reads is a complex task, primarily due to ambiguities in the reconstruction that are caused by genomic repeats [6]. In addition, metagenomic assemblers must tolerate the non-uniform representation of genomes in a sample as well as genomic variants between the sequences of closely related organisms. Despite advances in metagenomic assembly algorithms over the past years [6-10] the computational difficulty of the assembly process remains high and the quality of the resulting assemblies requires improvement.

Consequently, many analyses of metagenomic data are performed directly on unassembled reads [11-15], however the much shorter genomic context leads to lower accuracy [17].The need for effective and efficient metagenomic assembly approaches remains high, particularly since long read technologies (which partly mitigate the challenges posed by repeats [17-19]) are not yet effective in metagenomic applications due to lower throughput, higher costs, and higher required DNA quality and concentration [20-21].

Reference-guided, comparative assembly approaches have previously been used to assist the assembly of short reads when a closely related reference genome was available [22-23]. Such approaches work as follows: short sequencing reads are aligned to a reference genome of a closely related species, then their reconstruction into contigs is inferred from their relative locations in the reference genome [23]. This process overcomes, in part, the challenge posed by repeats as the entire read (not just the segment that overlaps within adjacent reads) provides information about its location in the genome.

To date, hundreds of thousands of bacterial genomes have been sequenced to a high level of quality [25], and this number is expected to grow rapidly thanks to long read technologies. These sequenced genomes provide a great resource for performing reference-guided assembly of metagenomic sequences. Techniques developed in the context of single genomes cannot, however, be directly used in a metagenomic setting. Simply mapping a set of reads to even hundreds of different genomes is currently computationally prohibitive. Furthermore, genome databases comprise many variants of a same genome (e.g., the US FDAs GenomeTrackr project [26] alone has contributed over 500,000 different strains of *Salmonella*), and genome-by-genome analyses would result in redundant reconstructions of metagenomic sequences. We also note that some recent reference-guided strategies implemented in genomic analysis tools, such as the "--trusted-contigs" feature of the SPAdes assembler [26-27] and StrainPhlan [29] ignore the fact that the data being reconstructed originates from genomes that are related but different from the genomes found in public databases. As a result, such approaches may mis-assemble the metagenomic data exactly within the genomic regions where novel biological signals may be located.

In this paper, we describe the first effective assembly software package for the reference-based assembly of metagenomic data. We rely on an indexing strategy to quickly construct sample-specific reference collections, thereby dramatically reducing the computational costs of mapping metagenomic reads to all references. We align reads against closely related genomes only once, then follow with a polishing step to resolve the discrepancies between the metagenomic data and the reference genomes. We show that our reference-based assembly approach yields high quality assemblies that generally outperform corresponding *de novo* assemblies of the same data, without introducing significant computational overhead.

Our software is released at https://gitlab.umiacs.umd.edu/mpop/metacompass under the BSD 3-Clause Clear License (https://choosealicense.com/licenses/bsd-3-clause-clear/).

# Results
## Data sources and overall statistics
We assessed MetaCompass's performance by analyzing 90 metagenomic samples from the Human Microbiome Project (HMP) [4], accessible at ftp://public-ftp.hmpdacc.org/Illumina/PHASEII/. These samples, 15 each from six different body sites—the tongue dorsum, buccal mucosa, posterior fornix, supragingival plaque, stool, and anterior nares—were randomly selected from the full HMP data set, with the intent of capturing microbial communities of varied diversity and sequencing depth (information about the samples is provided in Supplementary Table 1). To generate a baseline, we also generated *de novo* assemblies for all samples using metaSPAdes [30] (v3.15.5) and MEGAHIT (v1.2.9) [31], arguably the most commonly used assembly tools in metagenomic experiments.

The amount of sequencing data varied depending on body site, likely due to the different extent of human DNA contamination in each sample. Stool samples contained most sequencing reads (a median of 179.7M), followed by tongue dorsum (median 156.1M), supragingival plaque (median 117.6M), buccal mucosa (median 13.7M), posterior fornix (median 3M) and anterior nares (median 559K). Ten samples, nine from the anterior nares and one from buccal mucosa could not be assembled by MetaCompass because none of the reference genomes was covered at sufficient depth of coverage. The results shown below reference just the 80 samples that were assembled by MetaCompass. Detailed statistics for each individual sample can be found in the Supplementary Material file named "Supplementary_material.xlsx".

**Overall differences between MetaCompass and *de novo* assemblies**

We compared the overall output from MetaCompass with *de novo* assemblies of the same samples, generated using metaSpades. We excluded from analysis the 10 samples that could not be assembled by MetaCompass as well as an additional two samples that could not be assembled by metaSpades. The total size of the MetaCompass assemblies was lower than that produced by metaSpades, which was expected since MetaCompass can only assemble the fraction of the metagenome that aligns to reference genome sequences. Nonetheless, the MetaCompass assembly represented up to 97% of the metaSpades assembly for a posterior fornix sample, indicating that, in that sample, reference-guided assembly can be effectively used for the majority of the organisms in the sample. On average, the fraction of the *de novo* assembly that could be covered by the MetaCompass output ranged from 83% for posterior fornix to a low of 40% for the buccal mucosa (see Figure 1 for the distribution of these fractions across the samples).

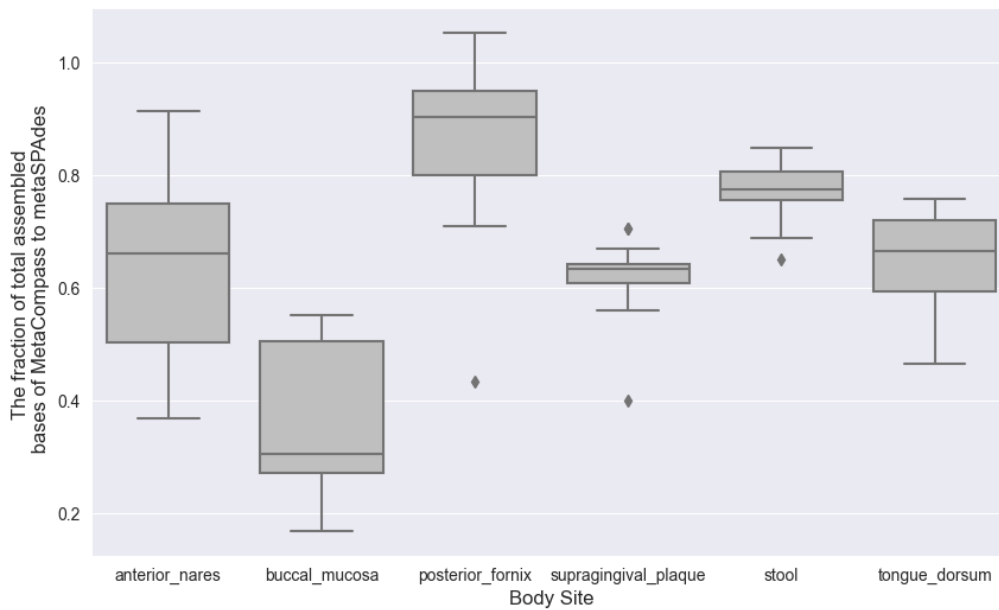

Figure 1 Comparing the total size of the MetaCompass assembly to that of metaSpades. Since MetaCompass can only assemble sequences that align to reference genomes, the total assembly size reflects the fraction of the sample "explained" by reference genome collections.

**Contiguity of MetaCompass assemblies of individual genomes**

To assess whether MetaCompass provides a benefit over *de novo* tools for the genomes selected as references, we compare the contiguity of the corresponding assemblies for each genome cluster. To evaluate contiguity, we use the metric NG25 [32], defined as follows. NG25 is the size of the largest contig *c* such that the sum of the lengths of contigs longer than *c* exceeds 25% of the entire genome size. This measure is only well-defined when the genome size is known, and we only apply it to assess the quality of the assembly within the context of individual reference genomes. In the context of a cluster of closely-related reference genomes, we use as a baseline the length of the longest genome in the cluster. Within each cluster, all contigs generated with the assistance of the references in the cluster are pooled together. We, then, use the same set of reads that was used by MetaCompass to generate *de novo* assemblies using metaSpades and MEGAHIT.

We focus our analysis on only the assemblies that cover a cumulative length of at least 25% of the reference genome's length. Figure 2 highlights the relationship between the NG25 and values and the depth of coverage, estimated with respect to the longest reference genome sequence within the corresponding cluster. For clarity, for each assembly we only highlight two values: the assembly that achieves the largest NG25 metric and the second-best assembly if MetaCompass is the "winner". If MetaCompass does not generate the most contigous assembly, it is reported as the second measure irrespective of its relative rank among the three tools.

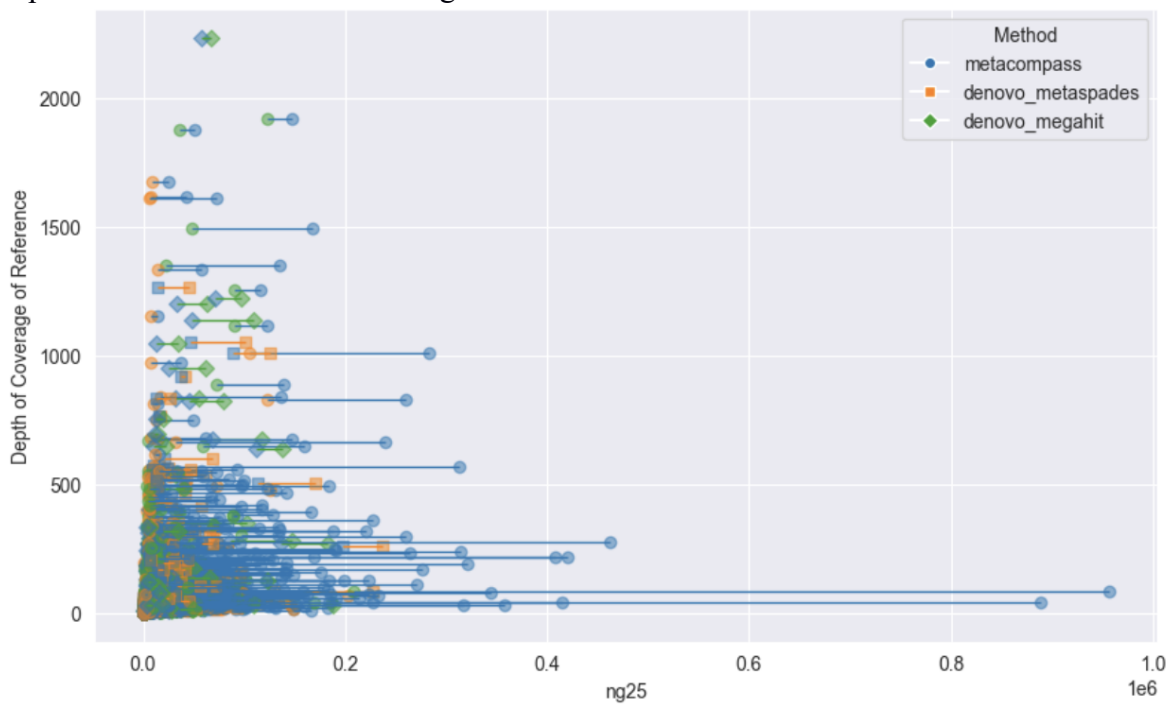

Figure 2. Comparison between MetaCompass and de novo assembly methods on the measurement of NG25 of the clusters vs the depth of coverage of references from all samples. The length of the line connecting the two assembly NG25 points of the sample cluster represents the difference between the NG values of the two points.

At smaller NGx values and lower coverage depths (lower left corner, x<50,000 y<250), the differences between MetaCompass and the two *de novo* assemblers are small and all methods are comparably likely to perform best for individual samples. Outside of this region, MetaCompass often outperforms the other two methods, evidenced by more prevalent blue lines in the figure. Moreover, when MetaCompass outperforms a *de novo* assembly tool, it frequently does so by a large extent, as demonstrated by the longer lines with the rightmost endpoint corresponding to MetaCompass.

Breaking down the analysis by body site (Figure 3) we note that MetaCompass typically outperforms the *de novo* tools with the exception of the posterior fornix where metaSPAdes has the best performance. The organisms where metaSPAdes showed the biggest benefit over MetaCompass were *Bifidobacterium breve*, *Lactobacillus iners*, *Lactobacillus jensenii*, *Lactobacillus gasseri*, and *Lactobacillus crispatus*, all important members of human vaginal microbiota. We also note that the overall contiguity of assemblies varied across body sites, with median NG25 values ranging from 10,182 for stool to 30,535 for posterior fornix. This contiguity variation trend was also observed in the *de novo* assemblies indicating fundamental characteristics of the sample impact contiguity more than specific algorithmic choices.

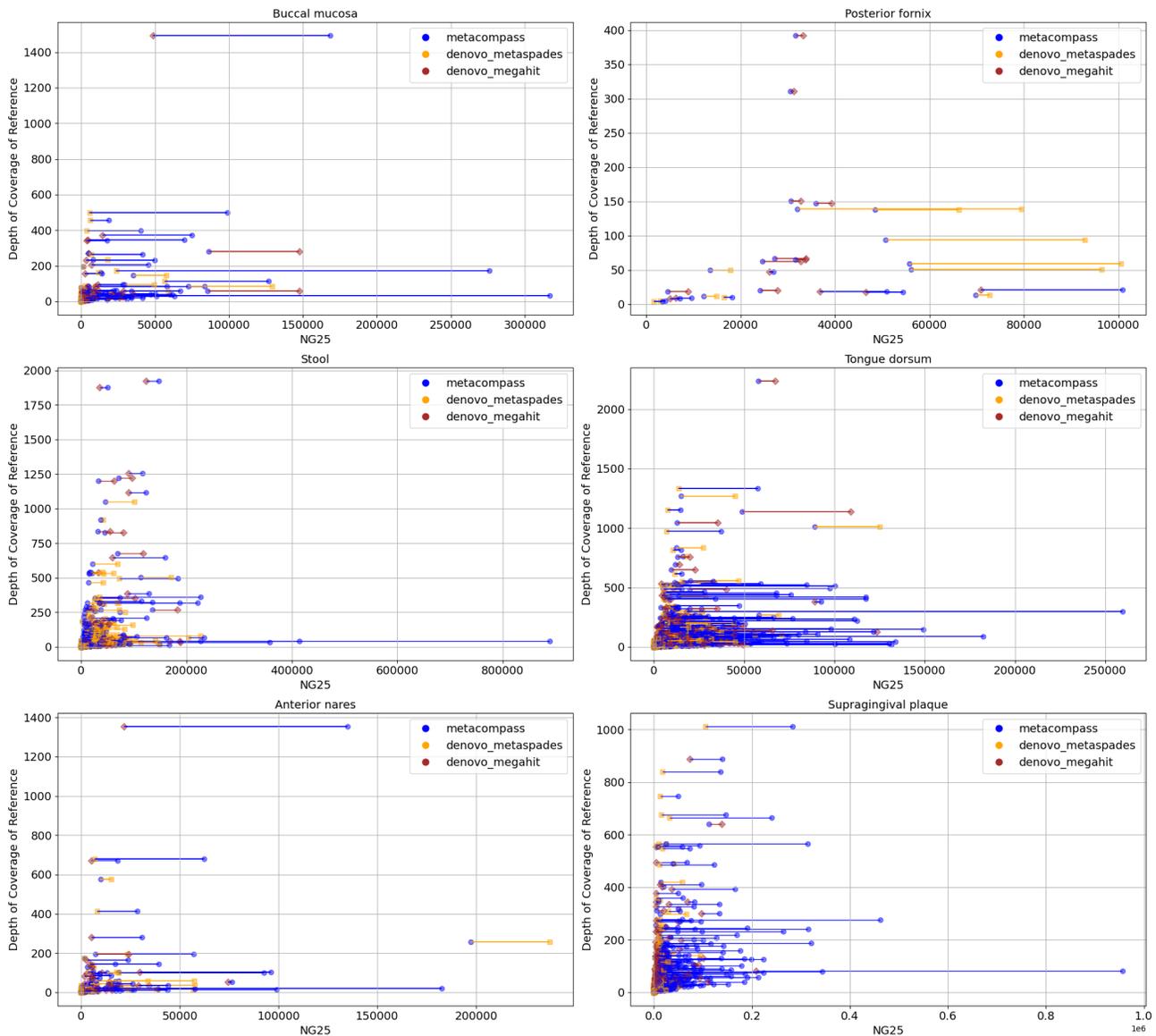

Figure 3 Comparison between MetaCompass and *de novo* assembly methods on the measurement of NG25 of the clusters versus the depth of coverage of references from 15 stool samples of each body site. The length of the line connecting the two assembly NG25 points represents the difference between the NG values of the two points.

## MetaCompass captures the pangenome of microbiome members

MetaCompass employs clustering as a key step in the analysis, grouping reference genomes into species-level clusters. Within these clusters, MetaCompass prioritizes reference genomes in descending order of k-mer similarity with the read set (as approximated by min-hashing distance). This strategy aims to ensure that the first genome assembled within a cluster is the one that is most similar to the corresponding genome within the sample. Reads that could not be aligned to the first reference genome are then iteratively aligned to the other reference genomes in the cluster, in decreasing order of their fit with the input data, with the intent of capturing genomic segments that were not present in the previously-assembled genomes. This functionality can only be leveraged if the reference database captures strain-level diversity, as measured here by the number of reference genome clusters that contain two or more genomes. The fraction of non-singleton clusters varies across body sites (Figure 4) with a high of 100% (all reference genome clusters comprise two or more reference genomes) for the posterior fornix to a low of 58.7% in supragingival plaques. We also note that for body anterior nares and buccal mucosa we observe a large extent of sample-to-sample variation in the ratio of non-singleton clusters captured.

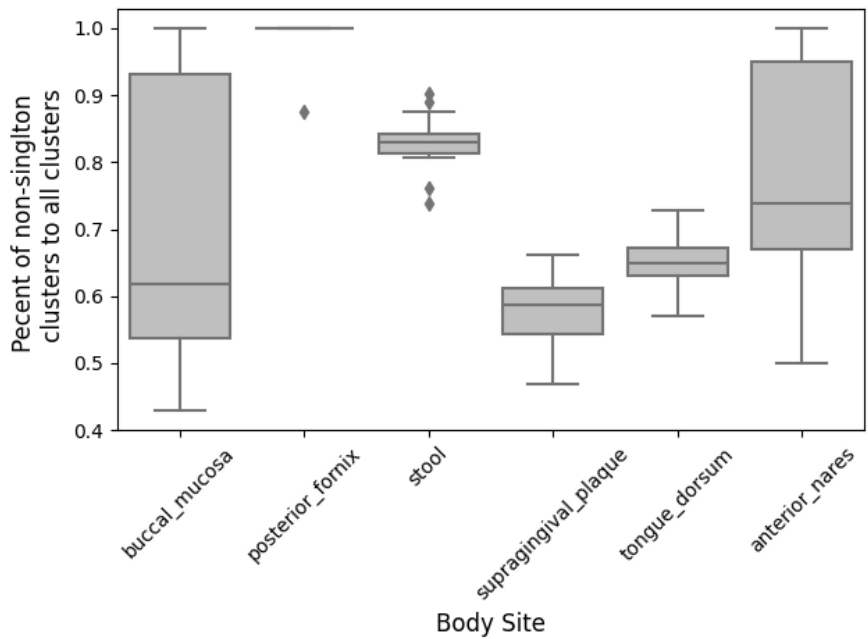

Figure 4 The distribution of non-singleton clusters across body sites.

Within multi-genome clusters, MetaCompass frequently uses more than one genome (Figure 5) demonstrating that the genomes found in the samples do not necessarily have a good fit with any specific reference genome. At one extreme, in stool samples, MetaCompass used a median of 6 genomes per cluster while in buccal mucosa and anterior nares, the median was close to 2.

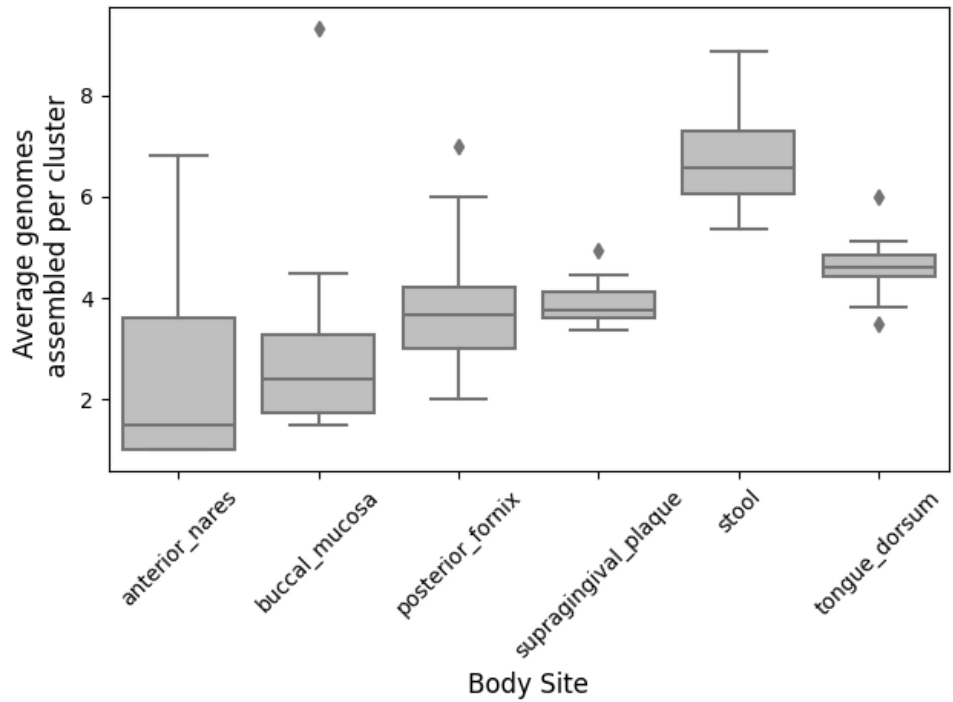

Figure 5. The distribution of the number of genomes assembled per cluster (excluding singleton clusters) across body sites.

To further demonstrate the use of multiple related reference sequences in the assembly process, we focus on cluster #11 from tongue dorsum sample SRR514250. This cluster includes three reference genome sequences from the species *Streptococcus infantis*. The genome sequence of strain *S. infantis* STn450 showed the highest k-mer similarity to the reads and was assembled first. MetaCompass effectively assembled 1,548,950 bases, representing 84.57% of the reference genome. It generated 173 contigs larger than 2,000 bases with the longest contig measuring 41,513 bases and with an NG25 of 16,098. The second strain selected, *S. infantis* ATCC 700779, resulted in an assembly covering 599,811 bases (32.30% of the reference genome), with 38 contigs greater than 2,000 bases, the largest contig of 8,457 bases and an NG25 of 625 bases. The assembly of the third genome, *S. infantis* NCTC13771, covered only 0.42% of the reference genome, with the longest contig and of 2,121 bases and a total of 8,531 bases assembled, leading MetaCompass to terminate processing this cluster.

**The effect of sequencing depth on the performance of MetaCompass**

To analyze the impact of depth of coverage on the effectiveness of MetaCompass, we focus on buccal mucosa sample SRR513142, which was the most deeply sequenced sample in our analysis comprising 13,723,918 reads. We sub-sampled the data to 80%, 60%, 40%, 20%, 10%, and 5% of its original input size and processed these data sets using MetaCompass.

At full coverage, MetaCompass identified 25 reference genome sequences forming 24 clusters. As coverage was reduced, we noted a steady decline in the number of reference sequences selected for assembly, the total bases assembled, the average breadth and depth of coverage and the number of marker genes covered in the reference selection process (see Table 1). At the 5% sampling rate, MetaCompass did not select any reference genomes, resulting in no assembly output.

Table 1. Average Depth of Coverage, Total Number of Marker Genes Covered, and Total References Assembled at Different Subsampling Rates for Buccal Mucosa Sample SRR513142

| Percentage of sub-sampling | Total number of read bases used by Metacompass | Number of total bases assembled | Average depth of coverage of the references (per base assembled) | Total number of marker genes covered in reference selection | Number of references assembled |
|---|---|---|---|---|---|
| 5% | 0 | 0 | 0 | 1,470 | 0 |
| 10% | 23,277,568 | 3,461,370 | 6.72 | 1,950 | 3 |
| 20% | 69,372,639 | 5,682,069 | 12.21 | 2,744 | 4 |
| 40% | 230,628,817 | 14,840,897 | 15.53 | 3,460 | 12 |
| 60% | 386,392,743 | 22,776,298 | 16.96 | 3,846 | 19 |
| 80% | 533,228,174 | 26,890,252 | 19.83 | 4,075 | 22 |
| 100% | 688,905,091 | 31,400,140 | 21.94 | 4,315 | 25 |

**MetaCompass achieves a high fraction of reads mapped.**

To assess how much of a sample is "explained" by mappings to reference genomes, we employ a metric termed "fraction of reads mapped", which was previously used in the literature to assess the "completeness" of metagenomic assemblies [30], [33].

$$Fraction\ of\ reads\ mapped\ =\ \frac{Total\ number\ of\ reads\ mapped\ in\ MetaCompass\ aseembly\ process}{Total\ number\ of\ reads\ in\ the\ sample}$$

Across the 80 samples that could be assembled by MetaCompass, the mean fraction of reads mapped was 74.2%, which is substantially better than the mapping rate obtained by manually selecting a collection of reference sequences (38.8%) reported in a study of genomic variation in the human gut microbiota [33].

The fraction of reads mapped varied across body sites and different initial numbers of input reads (Figure 5). Notably, samples from the buccal mucosa showed the lowest mean fraction of reads mapped of 51.67% followed by the anterior nares at 62.35%. These two body sites also exhibit a significantly larger variability of fraction of reads mapped across samples as compared to other body sites. The samples from all other sites achieved a mean fraction exceeding 73.49%.

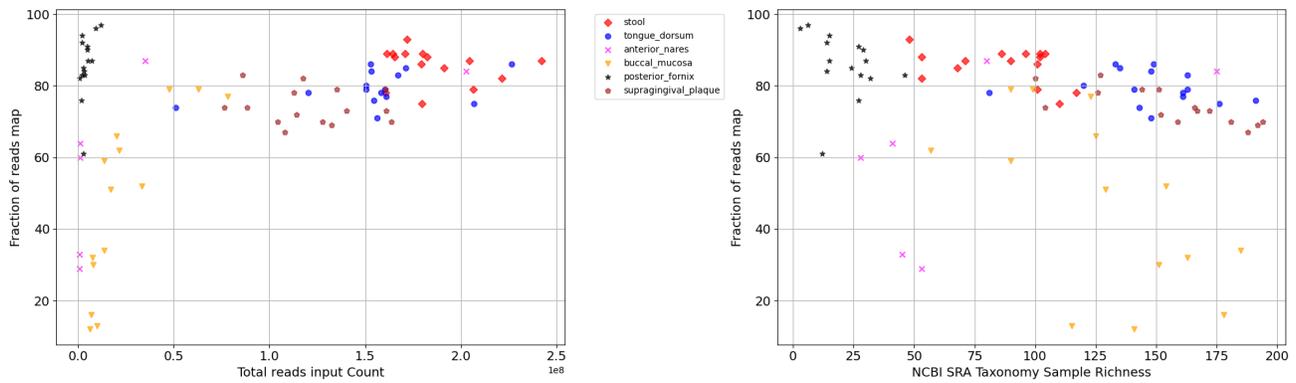

Figure 6. Dependence of read mapping rate and (left) total read count; and (right) sample richness. Broadly, read mapping rates increase with the total sequencing effort and decrease with sample richness. For the posterior fornix (black stars), high read mapping rates are achieved despite low depth of sequencing coverage due to the much lower richness of the samples.

We further explore the relationship between the size of the input, measured as the total number of reads in the sample, and the read utilization ratio (Figure 5). For the samples analyzed, MetaCompass achieved high read mapping rates for inputs with more than about 40,000,000 reads. The depth of sequencing needed reach a fraction of reads mapped exceeding 80% varied across different body sites. In the posterior fornix, high mapping rates could be achieved with as few as 4 million reads, while for the buccal mucosa the threshold could only be reached after 47 million reads. This variability among different body sites can be linked to sample complexity. Specifically, we rely on the sample richness as estimated from the NCBI SRA taxonomy analysis table (number of genomes at higher than 0.01% representation in the sample). The mean richness of posterior fornix samples is 24 (standard deviation: 11.39) while for the buccal mucosa samples it is 129 (standard deviation: 36.41).

Sequencing depth can also explain why certain samples could not be assembled by MetaCompass. The 9 anterior nares samples that failed to recruit reference genomes comprised a median of 463K reads, in contrast with a median of 1.0M reads for the samples that generated assemblies. The buccal mucosa sample that is not assembled has 3.6M reads whereas the other samples in this body site had a median of 15.4M reads.

### The effect of database size on assembly completeness and runtime

The default database used by MetaCompass includes 251,288 reference genomes from NCBI RefSeq, ensuring extensive coverage of bacterial diversity. To evaluate the impact of database size, we constructed an additional database from a much smaller collection of reference genomes, selecting only the genomes flagged as "representative" or "reference" from RefSeq. These typically represent one genome sequence per species that is determined by the RefSeq curators to best represent the species. After excluding the reference genomes that are missing genomic files, this database, referred to as the *reduced database*, includes 11,061 reference genome sequences. Note that using the reduced database eliminates the ability of MetaCompass to effectively sample the pangenome of a species.

As seen in Table 2 using the default database, MetaCompass consistently assembles more genomes than with the reduced database, a trend observed across a range of body sites and samples. For example, in the tongue dorsum sample SRR514250, we see 211 genomes assembled using the default database versus 36 with the representative database. Using the larger default database for assembly, however, leads to increased runtime. For instance, in the case of the tongue dorsum sample

SRR514250, the sample run on the default database was completed in over 16 hours in contrast to just 5 hours for the reduced database.

Table 2. Comparison of assembly statistics between MetaCompass with Default Database and Reduced Database, including number of reference genomes assembled and average number of references assembled per cluster.

| Body site | Sample | Number of reads | Sample Richness | # references default database | # references reduced database | Runtime default database | Runtime reduced database |
|---|---|---|---|---|---|---|---|
| tongue dorsum | SRR514250 | 226,602,332 | 133 | 211 | 36 | 16:29:48 | 05:24:49 |
| buccal mucosa | SRR513436 | 33,183,142 | 154 | 52 | 38 | 01:09:50 | 01:15:52 |
| posterior fornix | SRR513147 | 7,038,812 | 30 | 32 | 6 | 00:21:02 | 00:07:48 |
| posterior fornix | SRR628270 | 2,713,906 | 12 | 4 | 5 | 00:07:02 | 00:03:31 |
| supragingival plaque | SRR514844 | 112,950,078 | 126 | 115 | 45 | 06:50:27 | 03:16:16 |
| stool | SRR514206 | 171,825,210 | 48 | 104 | 14 | 07:56:40 | 02:37:06 |

**Computational Performance**

We evaluated the runtime performance of MetaCompass on a Linux 16-core server node (3.0GHz AMD® EPYC® 7313 Processor) and a memory ceiling of 256 GB. The evaluation was performed using a set of samples from different body sites and of varying sizes.

We assessed the runtime performance of MEGAHIT and metaSPAdes by processing the same samples with identical computing resources as those used for MetaCompass. The findings shown in Table 3 indicate that MetaCompass generally has longer runtime than MEGAHIT. However, its runtime is not significantly longer than that of metaSPAdes, except in the case of the tongue dorsum sample SRR514250. This particular sample exhibited much higher microbiome richness compared to other benchmarked samples and used a total of 221 reference genomes, in contrast to the second most diverse sample that only recruited 105 reference genomes.

Table 3. Assembly Statistics and runtime comparison of MetaCompass, MetaSPAdes, and MEGAHIT in HH:MM:SS

| Body site | Sample | Metacompass Runtime | metaSPAdes Runtime | MEGAHIT Runtime | Number of pair-end reads (in pairs) | Number of Metacompass reference genomes assembled |
|---|---|---|---|---|---|---|
| tongue dorsum | SRR514250 | 16:29:48 | 07:56:42 | 01:01:57 | 113,301,166 | 211 |
| buccal mucosa | SRR513436 | 01:09:50 | 01:19:10 | 00:16:48 | 16,591,571 | 52 |
| posterior fornix | SRR513147 | 00:21:02 | 00:14:47 | 00:02:00 | 3,519,406 | 32 |
| posterior fornix | SRR628270 | 00:07:02 | 00:05:26 | 00:00:50 | 1,356,953 | 4 |
| supragingival plaque | SRR514844 | 06:50:27 | 06:12:57 | 00:36:37 | 56,475,039 | 115 |
| stool | SRR514206 | 07:56:40 | 06:03:05 | 00:24:42 | 85,912,605 | 104 |

## Discussion

The goal of MetaCompass is to enable reference-guided approaches for sample analyses in order to leverage the substantial collection of genomic sequences currently available in public databases. As we demonstrate here, the MetaCompass assemblies of individual organisms within metagenomic samples generally outperform *de novo* assemblies of the same organisms. In other words, when appropriate reference genomes are available in the database, MetaCompass is an effective tool for reconstructing metagenome-assembled genomes (MAGs). It is important to note that reference-guided assembly of MAGs has advantages over *de novo* methods since the gene annotations and other information associated with the reference genome(s) used to guide the assembly are directly associated with the assembly without the need for further computation. Given the broad range of approaches used to construct MAGs through genome binning, we have only compared the results of MetaCompass to *de novo* assemblers, showing that the runtime of MetaCompass does not substantially exceed that of commonly used assembly tools. Since *de novo* assembly is just one of the steps used in *de novo* MAG construction, we believe that MetaCompass may match or improve upon the performance of MAG construction pipelines. Importantly, unlike many binning algorithms, MetaCompass can generate high-quality sequences without the need to analyze data from multiple samples.

The effectiveness of reference-based assembly, as implemented by MetaCompass, is significantly affected by the availability and relevance of reference genomes in the reference database being used. In vaginal samples we identified several genome clusters corresponding to important members of the human vaginal microbiota, where de novo assemblies were significantly better than the reference-guided assembly. This observation suggests that current databases do not adequately reflect the genomic diversity of the vaginal microbiota.

When comparing the assembly generated by MetaCompass on the basis of the complete reference database to that based on a reduced database of representative genome sequences, we noted that, for some samples, the reduced database yielded better assemblies. This observation is counter-intuitive because we would expect the broader genomic diversity in the entirety of RefSeq to better capture the genomic content of microbiome samples. Upon closer investigation we noted that in such situations, there was inconsistency between the estimated overlap between the reads and the reference genomes obtained through min-hashing and the actual usefulness of the reference genomes as a guide for assembly. In short, the order in which MetaCompass processed the genomes in a cluster was sub-optimal for these samples, suggesting the need for additional research towards developing fast methods for prioritizing the selection of reference genomes.

## Conclusion

We have described MetaCompass, a computational pipeline for reference-based metagenomic assembly. This novel method for metagenomic assembly leverages the increasing number of genome sequences available in public databases. Our findings demonstrate that reference-based assemblies provide advantages over *de novo* genome assemblies of the same organism. Downstream analysis, such as *de novo* assembly processes, can be conducted with the reads not utilized by MetaCompass to reconstruct the portion of the metagenomic sample that does not match known reference genomes. Future research could focus on the integration of MetaCompass and *de novo* assembly methods, for the complete assembly of the metagenomic samples.

## Methods

**Methods overview.** At a very high level, MetaCompass starts with a collection of genomes that could be used as references. In a reference selection step, we identify a subset of the reference collection representing genomes that could plausibly be used to guide the assembly of the sample being analyzed. We make this determination on the basis of the coverage of universal marker genes—genomes that have a large fraction of the marker genes sufficiently well covered by the reads in the samples are further considered. To address the large extent of redundancy in the reference collection, in a reference culling step we cluster the refined list of reference genomes based on average nucleotide identity (ANI). During the assembly stage, we proceed in a cluster-by-cluster basis, prioritizing clusters according to the k-mer similarity between each cluster and the read set. To assemble a single cluster, reads are aligned to the references, and contigs are extracted in the order of the extent of coverage of each reference. Consensus sequences are used by "polishing" the reference sequence using Pilon [34] (v1.18). Upon completion, MetaCompass provides assembly statistics and outputs all reads that were not utilized in the assembly process so that they are easily accessed for additional downstream analyses, such as de novo assembly. These individual steps are described in more detail below.

**Reference databases.** The default database consists of the high-quality genome sequences found in the NCBI RefSeq database. Specifically, these are genome sequences that are not flagged as "atypical" or "contaminated", or that are missing genomic CDS files. The database used to generate the results of this manuscript was constructed in March 2022, and comprised 251,288 genome sequences.

To evaluate the impact of the database size of the performance of MetaCompass, we also constructed a reduced database by selecting from RefSeq just those genomes that were marked as "reference" or "representative" in the database, i.e., genomes selected by the curators of the database due to their quality and ability to typify microbial species. In this database, we expect only one or a handful of genomes per species, leading MetaCompass to create mostly singleton clusters during the reference culling step. This database was constructed in August 2023 and comprised 11,061 genome sequences.

**Reference selection.** While comparative assembly approaches have already been described for single genomes [23] [32], their use in metagenomic data is complicated by the fact that the appropriate reference genome(s) need to be selected in a sample-specific manner from a potentially large collection of genomes (e.g., all publicly available microbial genome sequences). Building efficient indexes for large reference collections is computationally challenging for short-read aligners [35], both in terms of speed and memory consumption. Therefore, narrowing down the search space for potential reference genomes before initiating any whole genome indexing or read alignment process is crucial. To tackle this task, we leverage the fact that we are only interested in those genome sequences that will be useful to guide the assembly process, specifically, genome sequences that are sufficiently well covered by the reads in the sample. At a high enough depth of coverage (approx. > 3-fold), we can assume that the vast majority of genes in a genome is covered by reads. Thus, our initial indexing strategy focuses on a relatively small set of universal marker genes, retaining only those genomes that have the majority of these genes covered sufficiently well by the reads in the input.

We use FetchMG [15] to identify 40 universal single-copy marker genes across all reference genomes, in a one-time MetaCompass database-building process. Database indexing involves grouping sequences from each marker gene into clusters with a sequence similarity threshold of 99% to identify representative sequences for each cluster using CD-HIT [36] (version 4.8.1, using command "`cd-hit-est -c 0.99 -n 10 -d 0`"), and the index is reusable for future runs. During the execution of MetaCompass, all input reads are aligned using Minimap2 to the pre-extracted marker gene representative. For a reference genome to be considered for assembly, it must cover at least 75%of the universal marker genes, where a marker gene is considered covered by reads if the breadth of the alignment coverage exceeds 90% of the length of the gene. This step typically narrows down the total number of reference candidates for assembly from 251,288 to a few thousand.

**Reference culling.** The reference selection step described above selects a redundant set of genomes. For example, the many *Escherichia coli* genomes available in public databases would be selected if the sample being analyzed contains an *E. coli* strain. Adequately dealing with this ambiguity is critical for effective assembly. If all read mappings are retained, allowing a read to be associated with multiple reference genomes, the resulting assembly will be redundant, reconstructing multiple copies of the homologous genomic regions. If for each read a random placement is selected from among the multiple equivalent matches, none of the related genomes may recruit enough reads to allow assembly, thereby leading to a fragmented reconstruction. Assigning reads to genomes according to their estimated representation in the sample (determined, e.g., based on the number of reads uniquely mapped to each genome), may bias the reconstruction towards the more divergent reference genomes, which may lead to an overall poorer reconstruction of the genomic regions shared across related genomes.

To address these challenges, we developed a clustering-based approach. We cluster all reference genomes selected from previous steps based on average nucleotide identity (ANI) as estimated by Skani [37] (version 0.2.1, using command '`skani triangle`'), using a threshold of 95%—a widely-used cutoff for defining prokaryotic species boundaries [38]. MetaCompass prioritizes the assembly of the clusters on the basis of the overlap between the sequencing reads and the genomic sequences in the cluster. Reads assigned to one cluster are excluded from consideration in the subsequent clusters, thereby avoiding redundancy and reducing workload. The prioritization is accomplished by performing the intersection, using KMC [39] (v3.2.1, using command "`kmc -k28 -ci1 -hp -fq`"), between the unique k-mer set derived from the genome sequences in the cluster with the k-mers identified from the un-assigned reads. We iteratively process the clusters in order of the size of the kmer intersection with the reads, as re-computed before each cluster selection, and perform reference-guided assembly by using the reference genomes in the cluster in an iterative approach intended to further restrict the number of genomes processed.

**Cluster-based assembly.** Genomes within the same ANI-based cluster are closely related and have high genomic similarity, implying that reads are likely to map equally well to multiple reference sequences. To prevent redundancy, we process the genomes in the cluster in order of their overlap with the set of reads, and reads are eliminated from further consideration once used in the assembly of a genome. We use a strategy inspired by the greedy approximation algorithm for the set cover problem [40] by iteratively picking the reference genome to which we can align the majority of the unassigned reads. We are effectively trying to approximate finding the smallest number of reference genomes that "explains" the majority of the reads. The match between a reference genome sequence and the read set is approximated by the k-mer similarity between them using Mash [41](version 2.3, using the command '`mash screen -w -p 64`'). This approach is notably more efficient than direct read-to-genome alignment. Once the genome with the best match to the set of reads is identified, reads are aligned to it and assembled as described in the next section. The reads used in this assembly are removed from further consideration, and the process continues with the un-assembled reads. The assembly process continues until one of the following criteria is met: (i) all reads are assembled; (ii) all references in the cluster are assembled; (iii) the cumulative length of assembled contigs fails to reach 5% of the reference genome's length; or (iv) the longest contig assembled is shorter than 2,000 base pairs. The final two conditions detect when the process reaches a point of diminishing returns. When the cluster contains a single genome, the assembly is retained even if it fails conditions (iii) and (iv) above.

**Contig assembly.** The input reads are aligned to the currently-selected reference genome using Minimap2 [42] (version 2.26-r1175, using command `minimap2 -ax sr --heap-sort=yes`). The output of minimap2 is processed to identify regions of the reference that are not covered by any reads, and contigs are formed from the contiguous reference segments covered by reads. We remove from further consideration any contigs shorter than 500 base pairs. We record the locations of contigs within the reference genome and report it in an AGP-formatted file (format version 2.1) [43]. The contigs are then refined by polishing as described below.

**Contig polishing.** Our ultimate goal is to reconstruct the sequence of organisms found in the sample rather than simply recapitulating the sequence of the reference genome used as a guide. To account for differences between reads and the reference genome, we use Pilon[34], a widely used polishing tool. This process involves realigning all previously mapped reads to the contigs from the reference-guided assembly using Minimap2, then feeding the resulting BAM file into Pilon for polishing. In addition to the final consensus sequence generated by Pilon for each contig, the full report generated by Pilon is recorded as a part of MetaCompass output.

**MetaCompass output.** In addition to a FASTA-formatted file that contains the sequence of all the contigs, MetaCompass also reports the NCBI accessions of the references used in the assembly step. The relative placement of contigs along each reference is reported in an AGP-formatted file, together with the report generated by Pilon. Additionally, MetaCompass outputs all reads that were not used by MetaCompass, in fastq format, so that they can be fed into subsequent analysis steps, such as *de novo* assembly.

**Software availability**. MetaCompass is available as an open-source package at: https://gitlab.umiacs.umd.edu/mpop/metacompass. The code is licensed under the BSD 3-Clause Clear License: https://choosealicense.com/licenses/bsd-3-clause-clear/

**Data availability.** A total of 90 HMP samples had used for the analysis. A list of all samples used in this manuscript was provided as supplementary material in Sumplimentary_material.xlsx, accessible at https://obj.umiacs.umd.edu/metacompasssubmission/Supplementary_material.xlsx.

# List of abbreviations

**DACC** – Data Analysis and Coordination Center

**HMP** – Human Microbiome Project

**NCBI** – National Center for Biotechnology

**RefSeq** - NCBI Reference Sequence Database

**SRA** – Short read archive

# Declarations

**Ethics approval and consent to participate**
Not applicable.

**Consent for publication**
Not applicable.

**Competing interests**
The authors declare that they have no competing interests.

**Funding**
The authors were supported in part by the NIH, grants R01-HG-004885 and R01-AI-100947, by the NSF, grants IIS-1117247 and IIS-0812111, and the Office of Naval Research under cooperative agreement number N00173162C001, all to MP. SK was supported by the Intramural Research Program of the National Human Genome Research Institute, National Institutes of Health.

**Author's contributions**
TL, BL, and MP designed the approach. TL, VC, BL, ZB, UA, TT, and MA implemented the algorithms, methods, and scripts described in the paper. TL, VC, TT, and SK generated the assemblies presented in the paper. TL, TT, VC, and MA validated the assembly results and performed the comparisons between different assemblers. TL, TT, VC, MA, MP, and AP contributed to the analysis of the results. TL, VC, MP, BL, and TT wrote the paper. All authors were involved in reviewing and revising the manuscript. All authors read and approved the manuscript.